\documentclass[12pt]{article}
\usepackage{color,amssymb,epsfig,verbatim,amsmath}
\def\log{\hbox{log}}

\def\boxit#1{\vbox{\hrule\hbox{\vrule\kern6pt
          \vbox{\kern6pt#1\kern6pt}\kern6pt\vrule}\hrule}}
\def\refhg{\hangindent=20pt\hangafter=1}
\def\refmark{\par\vskip 2mm\noindent\refhg}

\def\refhg{\hangindent=20pt\hangafter=1}
\def\refmark{\par\vskip 2mm\noindent\refhg}

\def\bse{\begin{eqnarray*}}
\def\ese{\end{eqnarray*}}
\def\be{\begin{eqnarray}}
\def\ee{\end{eqnarray}}
\def\bq{\begin{equation}}
\def\eq{\end{equation}}
\def\bse{\begin{eqnarray*}}
\def\ese{\end{eqnarray*}}

\DeclareMathOperator*{\argmax}{argmax}

\usepackage{lscape}
\usepackage{pdflscape}
\usepackage{afterpage}

\begin{document}
\thispagestyle{empty}

\hfill\today \\ \\

\baselineskip=28pt
\begin{center}
{\LARGE{\bf  Variable selection and estimation for the additive hazards model subject to left-truncation, right-censoring and measurement error in covariates}}
\end{center}
\baselineskip=14pt
\vskip 2mm
\begin{center}
Li-Pang Chen\footnote{\baselineskip=10pt Corresponding Author: Department of Statistics and Actuarial Science, University 
of Waterloo, Waterloo, Ontario, Canada N2L 3G1, L358CHEN@uwaterloo.ca}

\end{center}
\bigskip

\vspace{8mm}

\begin{center}
{\Large{\bf Abstract }}
\end{center}
\baselineskip=17pt
{
High-dimensional sparse modeling with censored survival data is of great practical importance, and several methods have been proposed for variable selection based on different models. However, the impact of biased sample caused by left-truncation and covariates measurement error to variable selection is not fully explored. In this paper, we mainly focus on the additive hazards model and analyze the high-dimensional survival data subject to left-truncation and measurement error in covariates. We develop the three-stage procedure to correct the error effect, select variables, and estimate the parameters of interest simultaneously. Numerical studies are reported to assess the performance of the proposed methods.
}

\vspace{8mm}

\par\vfill\noindent

\par\medskip\noindent
\underline{\bf Short title}: Variable selection for additive hazards model

\clearpage\pagebreak\newpage
\pagenumbering{arabic}

\newlength{\gnat}
\setlength{\gnat}{22pt}
\baselineskip=\gnat

\clearpage

\section{Introduction} \label{Introduction}

Survival analysis has been proven useful in many areas including cancer research, clinical trials, epidemiological studies, actuarial science, and so on.
A large body of methods have been developed under various survival models and data subject to {\it right-censoring}. 
Comprehensive discussion on those methods can be found in Kalfleisch and Prentice (2002), Lawless (2003), and the references therein.
In practice, some complex features may appear in the dataset and make the analysis become challenging. In this paper, we mainly discuss \textit{left-truncation} and \textit{measurement error in covariates}. 

Left-truncation usually comes from the prevalent sampling design, in which individuals only experience the initiating event but not the failure event before the recruiting time. Under this sampling scheme, individuals might not be observed because they experience the failure event before the recruiting time. In this sense, left-truncation may cause the delayed entry of subjects and may tend to produce a biased sample. Several methods have been developed based on different types of models. For example, Qin and Shen (2010) proposed two different methods of the estimating equations to estimate $\beta$ based on Cox proportional hazards (PH) model. Huang, Follman, and Qin (2012) proposed the semiparametric likelihood inference for the Cox PH model based on the length-biased sampling which is a special case of left-truncation. Su and Wang (2012) developed the semi-parametric approach for the joint modelling between the left-truncated and right-censored survival outcomes and the longitudinal covariates. In addition, Shen, Ning, and Qin (2009) and Ning, Qin and Shen (2014) proposed valid methods to estimate the parameter for the accelerated failure time model.

Not only the models mentioned above, different type of models are also discussed in the developments of survival analysis based on specific purposes. For example, different from the investigation of the hazard ratio based on the Cox PH model, sometimes researchers may be more interested in the risk difference attributed to the risk factors. Based on this purpose, the additive hazards model is considered, and the formulation is given by 
\begin{equation} \label{additive_hazard_model}
\lambda(t|V) = \lambda_0 (t) + \beta^\top V,
\end{equation}
where $V$ is a $p$-dimensional vector of the covariates, $\lambda(t|V)$ is the conditional hazard function of the survival time given the covariates $V$, $\lambda_0(t)$ is the unspecified baseline hazard function, and $\beta$ is a $p$-dimensional vector of mainly interested parameter. Some methods have been proposed to deal with the additive model when left-truncation occurs. For example, Huang and Qin (2013) proposed the conditional estimating equation. Chen (2019a) developed the pseudo likelihood method to derive the estimator.

The second important feature is the measurement error in covariates. As discussed in Carroll et al. (2006), ignoring the error effect of covariates in the analysis may incur the tremendous bias of the estimator. With the absence of left-truncation, several methods have been developed to correct the error. To name a few, Nakamura (1992) developed an approximate corrected partial likelihood method which was extended by Buzas (1998) and Hu and Lin (2002). Huang and Wang (2000) proposed a nonparametric approach for settings with repeated measurements for mismeasured covariates. Xie et al. (2001) explored a least squares method to calibrate the induced hazard function. More related methods are also reviewed in Chen (2019c).

When both biased sample and measurement error occur simultaneously, several methods based on different type of models have been proposed. For example, Chen (2018) developed the three-stage procedure to deal with error-prone variables based on the accelerated failure time model. Chen (2019b) studied the cure model with left-truncated data and measurement error. Chen and Yi (2019) proposed the corrected pseudo likelihood estimation to estimate the parameter for the Cox PH model subject to left-truncated and right-censored survival data and covariate measurement error. However, other survival models, such as the additive hazards model, have not been fully explored when those two complex features occur in the dataset. Hence, in this paper, we mainly focus on the discussion of the additive hazards model.

On the other hand, high-dimensional data also attracts our attention. The analysis becomes difficult and the non-informative variables may appear as the dimension of variable increases. In order to collect the informative variables and make the analysis reasonable, the technique of \textit{variable selection} is one of useful tools to achieve this goal, and such method is also frequently implemented to the analysis of survival data. For example, Lin and Lv (2013) developed the variable selection method for the additive hazards model. However, they mainly focused on the survival data subject to right-censoring, and the analysis of variable selection with left-truncation and measurement error is not fully explored. 

In this paper, we consider this important problem and develop inference methods for analysis of high-dimensional left-truncated and right-censored survival data with measurement error. We mainly focus on the discussion of the additive hazards model. Different from the estimating equation approach, we adopt the pseudo likelihood method proposed by Chen (2019a), which provides the more efficient and robust estimator. Based on the pseudo likelihood method, we proposed the simulation-based three-stage procedure to correct measurement error, select the informative variables, and derive the estimators simultaneously. 

The motivated example of this paper is the Worcester Heart Attack Study (WHAS500) data which is collected by Hosmer et al. (2008).
The main goal of this study is to determine the factors associated with trends over time in the incidence and survival rates following hospital admission for acute myocardial infarction (MI). The data were collected over thirteen 1-year periods beginning in 1975 and extending through 2001 on all MI patients admitted to the hospitals in Worcester, Massachusetts. There are 500 observations and 22 variables in this dataset. Specifically, as discussed in Hosmer, Lemeshow, and May (2008), the beginning of survival time was defined as the time the subject was admitted to the hospital. The main interest is the survival time of a patient who was discharged and still alive. Hence, an inclusion criterion is that only those subjects who are discharged and still alive are eligible to be included in the analysis. That is, the minimum survival time would be the length of the time a patient stayed in the hospital; individuals whose observation times are shorter than the minimum survival time are not included in this analysis. 

Basically, the data are pertinent to three important events in calendar time:  time of hospital admission, time of hospital discharge, and time of last follow-up (which is either failure or censoring). The total length of follow-up is defined as the length of time between hospital admission and the last follow-up, and the length of hospital stay is defined as the time length between hospital admission and hospital discharge. Data can only be collected for those individuals whose total length of follow-up is longer than the length of hospital stay, which is the so-called left-truncation (e.g., Kalbfleisch and Prentice 2002, Section 1.3; Lawless 2003, Section 2.4).

The rest of this article is organized as follows. We first introduce the structure of left-truncated and right-censored (LTRC) survival data and measurement error model in Section~\ref{Data-structure}. We next present our proposed method in Section~\ref{Proposed-method}. Basically, we propose the three-stage procedure to deal with error-prone variables, select the active variables, and estimate the parameters simultaneously. In addition, we also provide the valid estimation procedure to derive the cumulative baseline hazards function and the distribution function of the truncation time. We give some model settings to examine the numerical performances of the estimator and 
implement the proposed method to WHAS500 dataset in in Section~\ref{Simulation}. Finally, the discussion of the paper is summarized in Section~\ref{Conclusion}.

\section{Notation and Model} \label{Data-structure}

\subsection{Data Introduction}
{
For an individual in the target disease population, let $\xi$ be the calendar time of the recruitment (e.g., the recruitment starts right at the hospital discharge) and let $u$ and $v$ denote the calendar time of the initiating event (e.g., hospital admission) and the failure event (e.g., death), respectively, where $u<v$ and $u < \xi < v$.  Let $T^\ast = v-u$ be the lifetime (e.g., the time length between the hospital admission and the failure) and $ A^\ast = \xi -u$ be the truncation time (e.g., the time length between the hospital admission and the hospital discharge). Let $V^\ast$ be a $p$-dimensional vector of covariates. Let $h(a)$ be the unspecified probability density function of $A^\ast$, and let $H(a) = \int_0^a h(\nu)d\nu$ denote the distribution function of $A^\ast$. Let $f(t)$ and $S(t)$ be the density function and the survivor function of failure time $T^\ast$, respectively.

Consistent with the notation considered by Chen (2019a, 2019b), for an individual with $T^\ast \geq A^\ast$, we let $\left(A,T,V\right)$ denote $\left(A^\ast,T^\ast,V^\ast\right)$ to indicate such an individual is eligible for the recruitment so that measuring $\left(A,T,V\right)$ is possible. Figure~\ref{fig:LTRC} gives an illustration of the relationship among those variables.
However, if $T^\ast < A^\ast$, as shown in Figure~\ref{fig:Truncation}, the individual is not included in the study so that the researcher cannot obtain any information of such individual. 

We further define $C$ as the censoring time for a recruited subject. Let $Y = \min \{ T, A+C \}$ be the observed time and let $\Delta = I(T \leq A+C)$ be the indicator of a failure event. Suppose we have a sample of $n$ subjects where for $i=1,\cdots,n$, $\left(Y_i,\Delta_i,A_i,V_i \right)$ has the same distribution of  $\left(Y,\Delta,A,V \right)$, and let $\left(y_i,\delta_i,a_i,v_i \right)$ denote the realization value.

For the following development, we make standard assumptions which are commonly considered for survival data analysis and related frameworks (e.g., Huang and Qin 2013; Chen 2019a):
\begin{enumerate}

\item[(A1)]Conditional on  $V^\ast$, $T^\ast$ are independent of $A^\ast$;

\item[(A2)] Censoring time is non-informative.
\end{enumerate}

\begin{figure}[!ht]
\centering
\begin{minipage}{1\textwidth}
\centering{\includegraphics[width=0.8\linewidth]{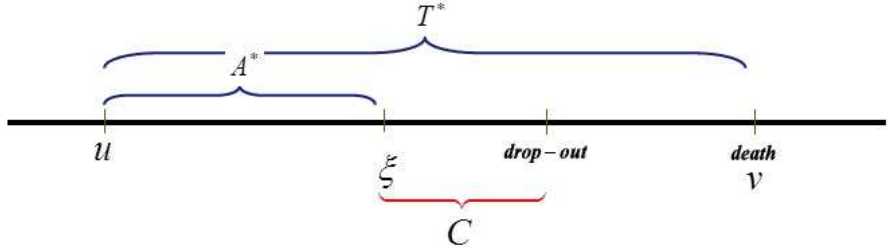}}
\end{minipage}
\caption{Schematic depiction of LTRC data for $T^\ast \geq A^\ast$} \label{fig:LTRC}
\end{figure}

\begin{figure}[!ht]
\centering
\begin{minipage}{0.9\textwidth}
\centering{\includegraphics[width=.8\linewidth]{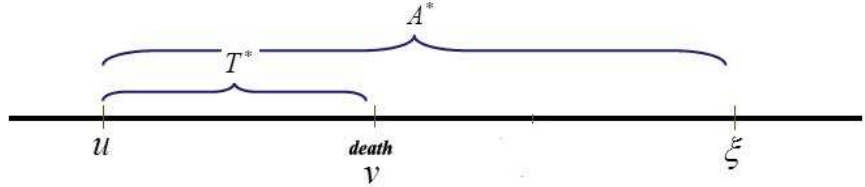}}
\end{minipage}
\caption{Schematic depiction of LTRC data. Truncation occurs when $T^\ast < A^\ast$} \label{fig:Truncation}
\end{figure}

}

\subsection{Measurement Error Model} \label{Def: Measurement}
 
In practice, covariates are often subject to measurement error. For $i=1,\cdots,n$, we write $V_i = \left( X_i^\top, Z_i^\top \right)^\top$, where $X_i$ and $Z_i$ are $p_x$-dimensional and $p_z$-dimensional vectors of the covariates, respectively. Moreover, we also decompose $\beta$ in (\ref{additive_hazard_model}) by $\beta = \left( \beta_x^\top, \beta_z^\top \right)^\top$, where $\beta_x$ and $\beta_z$ are $p_x$-dimensional and $p_z$-dimensional vectors of parameters associated with the covariates $X_i$ and $Z_i$, respectively. Let $p = p_x + p_z$.

 Suppose that $X_i$ is measured with error with an observed value or surrogate $W_i$, and that $Z_i$ is precisely observed. The classical additive measurement error model (Carroll et al. 2006, Ch1) is assumed to describe the relationship between $W_i$ and $X_i$:
\begin{equation} \label{mea_classic}
W_i = X_i + \epsilon_i,
\end{equation}
 where $\epsilon_i$ is independent of $\left\{ X_i,Z_i,C_i,A_i,T_i \right\}$, and $\epsilon_i \sim N(0,\Sigma_{\epsilon})$ with covariance matrix $\Sigma_{\epsilon}$.  
 If $\Sigma_\epsilon$ is unknown, then additional information, such as repeated measurements or validation data, is needed so that $\Sigma_\epsilon$ can be estimated. To ease of the discussion and focus the presentation on the analysis of impact on measurement error, we let $\Sigma_\epsilon$ be a known covariance matrix.

\section{Methodology} \label{Proposed-method}

In this section, we first briefly review the likelihood approach, which was proposed by Chen (2019a), based on the unobserved covariate $X$. After that, we present the extension by incorporating the error-prone variables and variable selection.

\subsection{Construction of the Likelihood Function} \label{Chen-2019}

Let $N_i(t) = \Delta_i I(Y_i \leq t)$ denote the counting process for the observed failure events. The modified  at-risk process is denoted by $R_i(t) = I(A_i \leq t \leq Y_i)$ for the adjustment of the truncation time.

Under Condition (A1), by the similar derivations in Appendix A of Chen (2019a), one can show that the joint density function of $(A,T)$ given $V = v$ is proportional to
\begin{equation} \label{joint_pdf}
\frac{f(t|v) dH(a)}{\int_0^\infty S(u|z)dH(u)} = \frac{f(t|v)}{S(a|v)} \times \frac{S(a|v) dH(a)}{\int_0^\infty S(u|z)dH(u)},
\end{equation}
where $\frac{f(t|v)}{S(a|v)}$ is the density function of $T$ given $A$ and $V$, and $\frac{S(a|v) h(a)}{\int_0^\infty S(u|v)h(u)du}$ is the density function of $A$ given $V$. Therefore, for $i=1,\cdots,n$, under Condition (A2) and model $(\ref{additive_hazard_model})$, the {\it full} likelihood function is given by 
\begin{equation} \label{Full}
L_F (\beta, \lambda_0,H) = \prod_{i=1}^n \frac{\left\{ \lambda_0 (y_i) + \beta_x^\top x_i + \beta_z^\top z_i \right\}^{\delta_i} S(y_i |x_i, z_i) dH(a_i)}{\int S(u|x_i,z_i) dH(u) },
\end{equation}
where $S(t |x_i, z_i) = \exp \left\{ -\Lambda_0 (t) - \left( \beta_x^\top x_i + \beta_z^\top z_i \right) t \right\}$ is the survivor function under model $(\ref{additive_hazard_model})$. 
Moreover, we can decompose $(\ref{Full})$ into $L_C \times L_M$ , where
\begin{equation} \label{Conditional}
L_C (\beta, \lambda_0) = \prod_{i=1}^n \frac{\left( \lambda_0 (y_i) + \beta_x^\top x_i + \beta_z^\top z_i \right)^{\delta_i} S(y_i | x_i, z_i)}{ S(a_i|x_i, z_i)}
\end{equation}
is the likelihood of $(Y,\Delta)$ given $A,X,Z$; and
\begin{equation} \label{Marginal}
L_M (\beta, \lambda_0,H) = \prod_{i=1}^n \frac{ S(a_i |x_i ,z_i) dH(a_i)}{\int S(u|x_i,z_i) dH(u)}
\end{equation}
is the likelihood of $A$ given $X$ and $Z$.

Different from the conventional martingale method or the estimating equation approach, Chen (2019a) derived the estimator of $\beta$ by maximizing the pseudo likelihood function based on (\ref{Full}). There are some advantages. The first advantage is that \textit{misspecification} is considered. That is, the property of the martingale method holds only if the model is correctly specified, while the likelihood method does not need such strong condition (Lin and Wei 1989). The second advantage is that the likelihood method gives the more robust and more efficient estimator. This property is shown by numerical studies in Chen (2019a).

\subsection{Inferential Procedure} \label{Proposed-Chen}

In this section, we extend the setting in Section~\ref{Chen-2019} by incorporating the error-prone and high-dimensional covariates. To deal with error-prone covariate, select active covariate variables, and estimate parameters simultaneously, we propose a simulation-based three-stage procedure.
\begin{description}
\item[Step 1:] {\it Simulation}

Let $B$ be a given positive integer and let $\mathcal{Z} = \left\{ \zeta_0, \zeta_1,\cdots, \zeta_M \right\}$ be a sequence of pre-specified values with
$0 = \zeta_0 < \zeta_1 < \cdots < \zeta_M$, where $M$ is a positive integer, and $\zeta_M$ is a prespecified positive number such as $\zeta_M = 2$. 

\text{\ \ \ \ }For a given subject $i$ with $i=1,\cdots,n$ and $b=1,\cdots,B$, we generate $U_{i,b}$ from $N(0,\Sigma_\epsilon)$, and define $W_i \left(b,\zeta\right)$ as
\begin{eqnarray} \label{conti_SIMEX}
W_i \left(b,\zeta\right) = W_i  + \sqrt{\zeta} U_{i,b}
\end{eqnarray}
for every $\zeta \in \mathcal{Z}$ and $b=1,\cdots,B$. Therefore, the conditional distribution of $W_i \left(b,\zeta\right)$ given $X_i$ is $N\left(X_i,(1+\zeta)\Sigma_\epsilon \right)$.
\item[Step 2:] {\it Estimation and selection}

We adopt the likelihood function in Section~\ref{Chen-2019}. Specifically, replacing $X_i$ by $W_i \left(b,\zeta\right)$ gives
\begin{equation} \label{Full-correct}
L_F^\ast (\beta, \lambda_0,H) = L_C^\ast (\beta, \lambda_0) \times L_M^\ast (\beta, \lambda_0,H),
\end{equation} 
where
\begin{equation} \label{Conditional-correct}
L_C^\ast (\beta, \lambda_0) = \prod_{i=1}^n \frac{\left( \lambda_0 (y_i) + \beta_x^\top w_i \left(b,\zeta\right) + \beta_z^\top z_i \right)^{\delta_i} S(y_i | w_i \left(b,\zeta\right), z_i)}{ S(a_i|w_i \left(b,\zeta\right), z_i)}
\end{equation}
 and
\begin{equation} \label{Marginal-correct}
L_M^\ast (\beta, \lambda_0,H) = \prod_{i=1}^n \frac{ S(a_i |w_i \left(b,\zeta\right) ,z_i) dH(a_i)}{\int S(u|w_i \left(b,\zeta\right),z_i) dH(u)}.
\end{equation}

\text{\ \ \ \ }By the similar derivations  in Chen (2019a), for given $b$ and $\zeta$, the estimators of $\Lambda_0(\cdot)$ and $\lambda_0(\cdot)$ are respectively determined by 
\begin{equation} \label{Baseline_Hazard-correct}
\widehat{\Lambda}_0 (t;\beta,b,\zeta) = \int_0^t \frac{\sum \limits_{i=1}^n \{ dN_i(u) - R_i (u) \left( \beta_x^\top W_i(b,\zeta) +  \beta_z^\top Z_i \right) du \}}{\sum \limits_{i=1}^n R_i(u)}
\end{equation}
and
\begin{equation} \label{baseline_kernel-correct}
\widehat{\lambda}_\sigma (y;b,\zeta) = \frac{1}{\sigma} \int K \left( \frac{y-\widetilde{y}}{\sigma} \right) d\widehat{\Lambda}_0 (\widetilde{y}; \beta, b,\zeta),
\end{equation}
where $\widetilde{y}$ is the independent copy of $y$, $K(\cdot)$ is the second order symmetric kernel function and  $\sigma$ is the positive-value bandwidth. The estimator of bandwidth $\sigma$ can be determined by the cross-validation criterion, and the detailed derivations can be found in Chen (2019a).

\text{\ \ \ \ }On the other hand, we observe that only (\ref{Marginal-correct}) involves $H(\cdot)$. To estimate it, it suffices to examine (\ref{Marginal-correct}). Different from the iteration method in Chen (2019a), here we use the nonparametric maximum likelihood estimator (NPMLE) (e.g., Wang 1991) to estimate the distribution function of $A^\ast$. For a fixed parameter $\beta$ and given $b$ and $\zeta$, the NPMLE of $H(a)$ in (\ref{Marginal-correct}) is given by
\begin{eqnarray} \label{est-H}
\widehat{H}(a;b,\zeta) = \left( \sum \limits_{i=1}^n \frac{1}{\widehat{S}(a_i | w_i(b,\zeta), z_i )} \right)^{-1} \sum \limits_{i=1}^n \frac{I(a_i \leq a)}{\widehat{S}(a_i | w_i(b,\zeta), z_i )},
\end{eqnarray}
where $\widehat{S}(a_i | w_i(b,\zeta), z_i ) = \exp\left\{ -\widehat{\Lambda}_0(a_i;\beta,b,\zeta) \exp\left(  \beta_x^\top w_i(b,\zeta) +  \widehat{\beta}_z^\top z_i \right) \right\}$ and $\widehat{\Lambda}_0(t;\beta,b,\zeta)$ is determined in (\ref{Baseline_Hazard-correct}).

\text{\ \ \ \ }Therefore, replacing the unspecified functions in (\ref{Conditional-correct}) and (\ref{Marginal-correct}) by  (\ref{Baseline_Hazard-correct}), (\ref{baseline_kernel-correct}), and (\ref{est-H}) gives the pseudo likelihood function $L_F^\ast (\beta, \widehat{\lambda}_0,\widehat{H})$, where $\widehat{\lambda}_0$ and $\widehat{H}$ represent (\ref{baseline_kernel-correct}) and (\ref{est-H}) for ease of notation.

\text{\ \ \ \ }To do the variable selection, we propose to use different penalty functions for $\beta$. Let $\rho(\beta)$ denote the penalty function and let $\vartheta$ be the tuning parameter. There are several choices of the penalty function, including the LASSO (Tibshirani 1996), adaptive LASSO (ALASSO, Zou 2006), and SCAD (Fan and Li 2001) methods. The detailed formulations are listed as follows:
\begin{itemize}
\item LASSO:

The penalty function based on the LASSO method is given by
\begin{eqnarray*}
\rho(\beta) = \sum \limits_{r=1}^p \left| \beta_r \right|.
\end{eqnarray*}
\item ALASSO:

The penalty function based on the ALASSO method is given by
\begin{eqnarray*}
\rho(\beta) = \sum \limits_{r=1}^p w_r\left| \beta_r \right|,
\end{eqnarray*}
where $w = \left( w_1,\cdots,w_p \right)$ is the vector of weights. As suggested by Zou (2006), the weight can be set as $w_r = \beta_r^{-\gamma_1}$ for any $\gamma_1 > 0$ and $r = 1,\cdots,p$. Noting that $\gamma_1 = 0$ gives $w_r = 1$ for all $r = 1,\cdots,p$, thus yielding the LASSO penalty. To find an estimate of $w_r$, one may first find a consistent estimate $\widetilde{\beta}$ of $\beta$ and then take $\widetilde{w}_r = \widetilde{\beta}_r^{-\gamma_1}$ as a weight for $r  = 1,\cdots,p$.
\item SCAD:

The penalty function based on the SCAD method is given by
\begin{eqnarray*}
\rho'(\beta) = I\left(\beta \leq \vartheta \right) + \frac{\left( a \vartheta - \beta \right)_+}{(a-1)\vartheta} I\left(\beta \geq \vartheta \right),
\end{eqnarray*}
where $(x)_+ = \max\{x,0\}$ and $a>0$ is a fixed parameter. As suggested by Fan and Li (2001), we let $a = 3.7$.
\end{itemize}
As a result, for the given $b$ and $\zeta$, we calculate
\begin{eqnarray} \label{est-beta-b-zeta}
\widehat{\beta}(b,\zeta) = \argmax \limits_{\beta} \left\{ L_F^\ast (\beta, \widehat{\lambda}_0,\widehat{H}) + \vartheta \rho(\beta) \right\}.
\end{eqnarray}

\text{\ \ \ \ }In implementing the proposed method, choosing sensible tuning parameters is critical. There is no unique way of selecting a suitable tuning parameter, and methods such as the Akaike information criterion (AIC), the Bayesian information criterion (BIC), the Cross Validation (CV), and the Generalized Cross Validation (GCV) may be considered. Suggested by Wang et al. (2007), BIC tends to outperform among those procedures, especially in the setting with a   penalized likelihood function. Consequently, we employ the BIC approach to select the tuning parameter $\vartheta$. To emphasize the dependence on the tuning parameter, we let $\widehat{\beta}(b,\zeta,\vartheta)$ denote the estimator obtained from (\ref{est-beta-b-zeta}). Define
\begin{eqnarray} \label{BIC_beta}
BIC_\beta(\vartheta) = 2nL_F^\ast\left(\beta, \widehat{\lambda}_0, \widehat{H}\right) + \log(n) \times \text{df} \left\{  \widehat{\beta}(b,\zeta,\vartheta) \right\},
\end{eqnarray}
where $\text{df} \left\{  \widehat{\beta}(b,\zeta,\vartheta) \right\}$ represents the number of non-zero elements in $ \widehat{\beta}(b,\zeta,\vartheta)$  for the given $\vartheta$. The optimal tuning parameter $\vartheta$, denoted by $\widehat{\vartheta}$, is determined by minimizing (\ref{BIC_beta}) within suitable ranges of $\vartheta$. As a result, the estimator of $\beta$ based on (\ref{est-beta-b-zeta}) is determined by $\widehat{\beta}(b,\zeta) = \widehat{\beta}(b,\zeta,\widehat{\vartheta})$.

\item[Step 3:] {\it Extrapolation}

Based on (\ref{est-beta-b-zeta}), we define
\begin{eqnarray*}
\widehat{\beta}(\zeta) =\frac{1}{B} \sum \limits_{b=1}^B \widehat{\beta}(b,\zeta)
\end{eqnarray*}
for any given $\zeta \in \mathcal{Z}$. For $r = 1,\cdots, p$, let $\widehat{\beta}_r(\zeta)$ denote the $r$th element of $\widehat{\beta}(\zeta)$. Then for each $r$ fit a regression model to the sequence $\left\{ \left(\zeta, \widehat{\beta}_r(\zeta)\right): \zeta \in \mathcal{Z} \right\}$ and extrapolate it to $\zeta = -1$. Let $\widehat{\beta}_r = \widehat{\beta}_r(-1)$ and denote $\widehat{\beta} = \left( \widehat{\beta}_1,\cdots,\widehat{\beta}_p \right)$ as the final estimator of $\beta$.

\end{description}

The key idea of the proposed three-stage procedure is to use simulated surrogate measurements to delineate the patterns of different degrees of measurement error on inference results. The first and third stages adopt the simulation-extrapolation (SIMEX) method (Cook and Stefanski 1994; Carroll et al. 2006, Chapter 5)  which is applicable to error-contaminated covariates. The second stage of the proposed method undertakes the selection of important variables for settings with different magnitudes of mismeasurement. It is imperative to address the impact of measurement error on variable selection in this step.

\subsection{Estimation of the Cumulative Baseline Hazards Function} \label{est-Lambda-0}

In this section, we discuss the procedure of estimating $\Lambda_0(\cdot)$ after the parameter $\beta$ is estimated in Sections~\ref{Proposed-Chen}.

Write $\widehat{\beta} = \left(\widehat{\beta}_x^\top, \widehat{\beta}_z^\top \right)^\top$. For $r = 1,\cdots,p_x$ and $s = 1,\cdots,p_z$, let $\widehat{\beta}_{x,r}$ denote the $r$th component in $\widehat{\beta}_x$ and let and $\widehat{\beta}_{z,s}$ be the $s$th component in $\widehat{\beta}_z$.
Let
\begin{eqnarray*}
\widehat{\mathcal{S}}_x = \left\{ r = 1,\cdots,p_x : \ \widehat{\beta}_{x,r} \neq 0 \right\}\ \text{and} \ \widehat{\mathcal{S}}_z = \left\{ s = 1,\cdots,p_z : \ \widehat{\beta}_{z,s} \neq 0 \right\}
\end{eqnarray*}
denote the sets containing the indices which reflect the non-zero components of the estimators $\widehat{\beta}_x$ and $\widehat{\beta}_z$, respectively. Moreover, define $\widehat{\mathcal{S}} = \widehat{\mathcal{S}}_x \cup \widehat{\mathcal{S}}_z$.
 Let $\widehat{\beta}_{\widehat{\mathcal{S}}}$ denote the subvector of $\widehat{\beta}$ containing non-zero elements based on $\widehat{\mathcal{S}}$. In addition, let $W_{i,\widehat{\mathcal{S}}_x}(b,\zeta)$ and $Z_{i,\widehat{\mathcal{S}}_z}$ denote two subvectors of $W_i(b,\zeta)$ and $Z_i$ containing non-zero elements based on $\widehat{\mathcal{S}}_x$ and $\widehat{\mathcal{S}}_z$, respectively.
For $b = 1,\cdots,B$ and $\zeta \in \mathcal{Z}$, replacing $\beta$ by $\widehat{\beta}_{\widehat{\mathcal{S}}}$ in (\ref{Baseline_Hazard-correct}) gives
\begin{equation} \label{est-Lambda-b-zeta}
\widehat{\Lambda}_{\widehat{\mathcal{S}},0} (t;b,\zeta) = \int_0^t \frac{\sum \limits_{i=1}^n \{ dN_i(u) - R_i (u)  \widehat{\beta}_{\widehat{\mathcal{S}}}^\top \left( W_{i,\widehat{\mathcal{S}}_x}^\top(b,\zeta), Z_{i,\widehat{\mathcal{S}}_z}^\top \right)^\top du \}}{\sum \limits_{i=1}^n R_i(u)}
\end{equation}
for a given time $t$. Taking averaging on (\ref{est-Lambda-b-zeta}) with respect to $b$ gives
\begin{eqnarray} \label{est-Lambda-zeta}
\widehat{\Lambda}_{\widehat{\mathcal{S}},0} (t;\zeta) =  \frac{1}{B} \sum \limits_{b=1}^B \widehat{\Lambda}_{\widehat{\mathcal{S}},0} (t;b,\zeta) \ \ \text{for}\ \ \zeta \in \mathcal{Z},
\end{eqnarray}
where $t$ is a given time.

To estimate the cumulative baseline hazard function $\Lambda_0(t)$ at a given time point $t >0$, we adopt Step 3 in Section~\ref{Proposed-Chen} and fit a regression model to $\left\{ \left( \zeta, \widehat{\Lambda}_{\widehat{\mathcal{S}},0} (t;\zeta) \right) : \zeta \in \mathcal{Z} \right\}$ through a regression function $\varphi_\Lambda(\zeta;\Gamma_\Lambda)$ with the associated parameter denoted by $\Gamma_\Lambda$, i.e.,
\begin{eqnarray} \label{reg-Lambda-0}
\widehat{\Lambda}_{\widehat{\mathcal{S}},0} (t;\mathcal{Z}) = \varphi_\Lambda(\mathcal{Z};\Gamma_\Lambda) + \eta_\Lambda
\end{eqnarray}
with a noise term $\eta_\Lambda$, then we extrapolate it to $\zeta = -1$. The resulting value, denoted as $\widehat{\Lambda}_{\widehat{S},0} (t)$, is taken an estimate of $\Lambda_0(t)$.

\subsection{Estimation of the Distribution Function of Truncation Time} \label{est-H-0}

Once $\widehat{\beta}$ is obtained, we can also derive the estimator of $H(\cdot)$, and the procedure is parallel with the idea in Section~\ref{est-Lambda-0}. Specifically, first replacing $\beta$ in (\ref{est-H}) by  $\widehat{\beta}$ based on $\widehat{\mathcal{S}}$ gives
\begin{eqnarray} \label{est-H-b-zeta}
\widehat{H}_{\widehat{\mathcal{S}}}(a;b,\zeta) = \left( \sum \limits_{i=1}^n \frac{1}{\widehat{S}_{\widehat{\mathcal{S}}}(a_i | w_i(b,\zeta), z_i )} \right)^{-1} \sum \limits_{i=1}^n \frac{I(a_i \leq a)}{\widehat{S}_{\widehat{\mathcal{S}}}(a_i | w_i(b,\zeta), z_i )}
\end{eqnarray}
for a given time $a$, where $\widehat{S}_{\widehat{\mathcal{S}}}(a | w_i(b,\zeta), z_i ) = \exp\left[ -\widehat{\Lambda}_{\widehat{\mathcal{S}},0} (a;b,\zeta) \exp\left\{  \widehat{\beta}_{\widehat{\mathcal{S}}}^\top \left( W_{i,\widehat{\mathcal{S}}_x}^\top(b,\zeta), Z_{i,\widehat{\mathcal{S}}_z}^\top \right)^\top \right\} \right]$ and $\widehat{\Lambda}_{\widehat{\mathcal{S}},0} (t;b,\zeta)$ is determined in (\ref{est-Lambda-b-zeta}).

Next, taking average on (\ref{est-H-b-zeta}) with respect to $b$ gives
\begin{eqnarray} \label{est-H-zeta}
\widehat{H}_{\widehat{\mathcal{S}}}(a;\zeta) = \frac{1}{B} \sum \limits_{b=1}^B \widehat{H}_{\widehat{\mathcal{S}}}(a;b,\zeta)
\end{eqnarray}
for $\zeta \in \mathcal{Z}$, where $a$ is a given time.

Finally, similar to (\ref{reg-Lambda-0}), we fit a regression model to $\left\{\left(\zeta, \widehat{H}_{\widehat{\mathcal{S}}}(a;\zeta) \right) : \zeta \in \mathcal{Z} \right\}$ and then extrapolate it to $\zeta = -1$. Consequently, the resulting value, denoted as $\widehat{H}_{\widehat{\mathcal{S}}}(a)$, is taken an estimate of $H(\cdot)$.

\section{Numerical Studies} \label{Simulation}

\subsection{Model Setting}

Let $n$ be the sample size and here we keep $n = 400$. Let $\beta_{x0} \in \mathbb{R}^{p_x}$ and $\beta_{z0} \in \mathbb{R}^{p_z}$ be the true parameters as described in (\ref{additive_hazard_model}), and we denote $\beta_{0} = \left( \beta_{x0}^\top, \beta_{z0}^\top \right)^\top$. Here we let $p = p_x + p_z$ and $p_x = p_z$. We consider $p_x = p_z = 15$ or $20$, which indicates $p=30$ or $40$. Let $\mathcal{S} = \{ r : \beta_r \neq 0, r = 1,\cdots,p \}$ denote the set containing non-zero elements, and $q = |\mathcal{S}|$ is the number of elements in $\mathcal{S}$. For the entries of $\beta_{x0}$ and $\beta_{z0}$, we let $\beta_{x0} = \beta_{z0} = \left( \underbrace{1,\cdots,1}_{\left[\frac{p_x}{4}\right]}, \underbrace{-1,\cdots,-1}_{\left[\frac{p_x}{4}\right]}, \underbrace{0,\cdots0}_{1-2\left[\frac{p_x}{4}\right]}, \right)$, where $[\cdot]$ stands for the Gauss integer.

Let $\Sigma = \left(
\begin{array}{c c}
\Sigma_x & \Sigma_{xz} \\
\Sigma_{xz}^\top & \Sigma_z 
\end{array}
\right)$, where $\Sigma_{xz}$ is the $p_x \times p_z$ covariance matrix of $X$ and $Z$ with entries $\sigma_{xzij}$, $\Sigma_x$ and $\Sigma_z$ are, respectively, $p_x \times p_x$ and $p_z \times p_z$ covariance matrices with entries $\sigma_{xij}$ and $\sigma_{zij}$ for $i,j=1,\cdots,p_x$. In particular, we let $\sigma_{xzij} = 0.5^{(2+|i-j|)}$, $\sigma_{xij} = \sigma_x^2 \rho_x^{|i-j|}$ and $\sigma_{zij} = \sigma_z^2 \rho_z^{|i-j|}$ with $\sigma_x^2 = \sigma_z^2 = 1.0$ and $\rho_x = \rho_z = 0.6$ for $i,j=1,\cdots,p_x$. Therefore,  let the covariates $(X^\ast{}^\top, Z^\ast{}^\top)^\top$ be generated by normal distribution $N\left(\mathbf{0}_{p}, \Sigma
\right)$, where $\textbf{0}_{p}$ is the $p$-dimensional zero vector.

Four model formulations for $(A^\ast,T^\ast)$ are considered in this simulation study as follows:
\begin{description}
\item[Model 1:] $\lambda(t|X^\ast,Z^\ast) = 0.5 \sqrt{t} + \beta_{x0}^\top X^\ast + \beta_{z0}^\top Z^\ast$, $A^\ast \sim U(0,100)$;      
\item[Model 2:] $\lambda(t|X^\ast,Z^\ast) = 0.5 \sqrt{t} + \beta_{x0}^\top X^\ast + \beta_{z0}^\top Z^\ast$, $A^\ast \sim \exp(10)$;   
\item[Model 3:] $\lambda(t|X^\ast,Z^\ast) =  \log(t) + \beta_{x0}^\top X^\ast + \beta_{z0}^\top Z^\ast$, $A^\ast \sim \exp(10)$;
\item[Model 4:] $\lambda(t|X^\ast,Z^\ast) = \exp(2t) + \beta_{x0}^\top X^\ast + \beta_{z0}^\top Z^\ast$, $A^\ast \sim \exp(10)$.    
\end{description}
  The observed data $(A,T,X,Z)$ is collected from $(A^\ast,T^\ast,X^\ast, Z^\ast)$ by conditioning on that $T^\ast \geq A^\ast$. We repeatedly generate data these steps we obtain a sample of a required size $n=400$. For the measurement error process, we consider model (\ref{mea_classic})  with error $\epsilon \sim N \left( 0, \Sigma_\epsilon \right)$, where $\Sigma_{\epsilon}$ is the diagonal matrix where the diagonal entry is taken as $ 0.01$, $0.5$, or $0.75$.

Let $C$ be the censoring time generated from the uniform distribution $U(0,c)$, where $c$ is a constant that is chosen to yield about 50\% censoring rate.
Consequently, $Y$ and $\Delta$ are determined by $Y = \min \left\{ T, A+C \right\}$ and $\Delta = I \left( T \leq A + C \right)$, and the sample with size $n = 400$ is $\left\{ (Y_i, \Delta_i, A_i, W_i, Z_i) \right\}$.

In implementing the proposed method, we set $B = 500$ and partition the interval $[0,2]$ into subintervals with the equal width 0.25 with the resulting cutpoints set as the values of $\zeta$. We take the regression function in Step 3 of the proposed method to be the quadratic polynomial functions, as suggested in Carroll et al. (2006, p.126). 

Finally, we perform  1000 simulations for each setting.

\subsection{Simulation Results}

To assess the performance of the estimator of $\beta$, we report several measures, the $L_1$-norm
\begin{eqnarray*}
\left\| \Delta_\beta \right\|_1 = \sum \limits_{i} \left| \widehat{\beta}_i - \beta_{0,i} \right|
\end{eqnarray*}
and the $L_2$-norm
\begin{eqnarray*}
\left\| \Delta_\beta \right\|_2 = \sqrt{ \sum \limits_{i} \left( \widehat{\beta}_i - \beta_{0,i} \right)^2},
\end{eqnarray*}
where $\Delta_\beta = \widehat{\beta} - \beta_0$. In addition, we calculate the number of the correctly selected variables (\#S) and the number of the falsely excluded variables (\#FN).

For Models 1-4, we compare the performance of the estimators obtained from applying the proposed method to the surrogate covariates as opposed to the estimators obtained from fitting the data with the true covariate measurements. We examine three different penalty functions as discussed in Section~\ref{Proposed-Chen}, including the LASSO, ALASSO, and SCAD methods. In comparison, we also examine the \textit{naive estimators} of $\beta$, denoted by $\widehat{\beta}_\text{naive}$, which is derived by directly implementing the observed covariates $W_i$ in (\ref{Full}).

In Tables~\ref{Simulation_A}-\ref{Simulation_D}, we report the numerical results of our proposed method and the naive approach as well as those obtained from the true covariate measurements. It is clear and expected that the results obtained from using the true covariate measurements are the best with the smallest norms under all settings. Regarding the performance on the proposed method with the three different penalty functions, the ALASSO and SCAD tend to slightly outperform the LASSO in terms of the specificity and the finite sample biases, indicated by the $L_1$-norm and $L_2$-norm. In terms of correctly selecting variables, the LASSO method includes more variables than  the ALASSO and SCAD methods. All methods perform equally well in terms of falsely excluding variables and sensitivity, producing nearly perfect results. Furthermore, it is revealed that the naive method performs unsatisfactorily, with considerable finite sample biases produced and unreliable variable selection and exclusion results.

\subsection{Analysis of the Worcester Heart Attack Study (WHAS500) Data}

In this section, we apply the proposed method to analyze the data arising from the Worcester Heart Attack Study (WHAS500), which is described in Section~\ref{Introduction}. Specifically, as discussed by Hosmer, Lemeshow, and May (2008), the beginning of a survival time was defined as the time that subject was admitted to a hospital. The main interest is in the survival times of patients who were discharged alive from hospitals. Hence, a selection criterion was imposed that only those subjects who were discharged alive were eligible to be included in the analysis. That is, their minimum survival time would be the length of their hospital stay; individuals whose failure times did not exceed the minimum survival time were not enrolled in this analysis, and hence the left-truncation happens. With such a criterion, a sample of size 461 was selected and the truncation rate was approximately 7.8\%. Be more specifically, total length of follow-up (lenfol) is the last event time (i.e., $Y_i = \min \left\{T_i,A_i + C_i \right\}$), length of hospital stay (los) is the truncation time (i.e., $A_i$), and vital status at last follow-up (fstat) is $\delta_i$. These 461 patients contribute the measurements which satisfy the constraint $T_i \geq A_i$. In this dataset, the censoring rate is 61.8\%.

The following covariates are included in our analysis: initial heart rate (\texttt{hr}, $X_1$), initial systolic blood pressure (\texttt{sysbp}, $X_2$), initial diastolic blood pressure (\texttt{diasbp}, $X_3$), body mass index (\texttt{bmi}, $X_4$), history of cardiovascular disease (\texttt{cvd}, $Z_1$),  atrial fibrillation (\texttt{afb}, $Z_2$), cardiogenic shock (\texttt{sho}, $Z_3$), age at hospital admission (\texttt{age}, $Z_4$), 
gender (\texttt{gender}, $Z_5$), congestive heart complications (\texttt{chf}, $Z_6$), complete heart block (\texttt{av3}, $Z_7$), MI Order (\texttt{miord}, $Z_8$), and MI Type  (\texttt{mitype}, $Z_9$). As indicated by  Bauldry et al. (2015) and Rothman (2008), it is reasonable to assume that covariates $X_1$, $X_2$, $X_3$ and $X_4$ are subject to mismeasurement due to the reasons including inaccurate measurement devices and/or procedures, the biological variability, and temporal variations. In this dataset, we have $p_x = 4$ and $p_z = 9$, yielding that $p = p_x + p_z = 13$.

Since this dataset contains no additional information, such as repeated measurements or validation data, for the characterization of the measurement error process, we conduct sensitivity analyses to investigate the measurement error effects.  Specifically, let $\Sigma$ be the sample covariance matrix of $\left(X_1,X_2,X_3,X_4 \right)^\top$, and for sensitivity analyses we consider $\Sigma + \Sigma_e$ to be the covariance matrix $\Sigma_\epsilon$ for the measurement error model (\ref{mea_classic}), where $\Sigma_e$ is the diagonal matrix with diagonal elements $\sigma_e$, which is specified as $\sigma_e = 0.15$, $0.5$, or $0.75$. 
  
Table~\ref{RDA_A} summarizes the estimates  the result of variable selection of both the proposed and the naive methods. We first observe that the LASSO method produces more variables than the SCAD and ALASSO methods. The proposed method with three different penalty functions gives the robust result of variable selection regardless of the degrees of error effect. It is interesting to see that   both ALASSO and SCAD methods select the same variables regardless of values of $\sigma_e^2$, it indicates that ALASSO and SCAD are highly recommended to adopt in analysis. Compared with the proposed method, we can observe that the naive method selects more variables. In addition, there are several variables which are commonly selected based on both methods, including \texttt{sysbp}, \texttt{diasbp}, \texttt{bmi}, \texttt{afb}, \texttt{av3}, \texttt{miord}, and \texttt{mitype}.

\section{Conclusion} \label{Conclusion}

Variable selection and estimation for survival data are always important topics and also attract our attentions. Even though several methods have been proposed to deal with these problems, there has been little work of addressing these two complex features simultaneously in inferential procedures. In this paper, we develop the three-stage procedure to simultaneously correct error-prone variables, select variables, and estimate the parameters of main interest. We further demonstrate satisfactory finite sample performance of our methods using simulation studies.

One of the advantage is that the proposed method is based on the pseudo likelihood approach, which produces the more robust and efficient estimator (Chen 2019a). In addition, the proposed three-stage procedure can be naturally extended to other models. It implies that the proposed method provides a flexible approach to deal with different situations.

Finally, even though we have developed the valid method with such complex setting, there are still some challenges and extensions. For example, mismeasurement in the discrete covariates may happen, and it is also called \textit{misclassification} problem. It is also interesting to explore misclassification, even mixture of measurement error and misclassification. In addition, even we discuss the high-dimensional data analysis, but we only consider the case $p <n$. Actually, the \textit{ultrahigh-dimensional} data analysis, i.e., $p \gg n$, is also an important topic. Finally, even though we have no theoretical results of the proposed method in the current manuscript, numerical results provide the satisfactory performance of the proposed method, including precise estimation and high accuracy of variable selection. Exploring theoretical results of the proposed method, including consistency or oracle property, is also an important work in the future.

\section*{Conflict of Interest}

This paper has no Conflict of Interest.

\clearpage

\begin{table}
       \huge
     \caption{Simulation result for Model 1 }\label{Simulation_A}

   \scriptsize

 \centering
  \renewcommand{\arraystretch}{0.9}
 \begin{tabular}{c c c c  c ccccc cccc ccc}

 \\
 \hline\hline
$p\ (q)$ & $\sigma_{\epsilon}$  & Method & \multicolumn{4}{c} { Result of proposed estimator $\widehat{\beta}$  } & & \multicolumn{4}{c} {Result of naive estimator $\widehat{\beta}_\text{naive}$ } \\ \cline{4-7} \cline{9-12}
 & & & $L_1$-norm & $L_2$-norm & \#S & \#FN
 & &  $L_1$-norm & $L_2$-norm & \#S & \#FN \\ 
 \hline 
30 (17) & 0.15 & LASSO & 1.290 & 0.101 & 18.450 & 0.000 & & 5.241 & 0.608 & 23.450 & 0.001 \\
    &      & SCAD  & 1.160 & 0.094 & 17.926 & 0.000 & & 4.179 & 0.355 & 20.926 & 0.006 \\
    &      & ALASSO  & 1.150 & 0.094 &  17.965 & 0.000 & & 4.506 & 0.424 &  21.994 & 0.003 \\
    \\
   & 0.50 & LASSO & 1.481 & 0.149 & 18.822 & 0.000 & & 7.178 & 0.759 & 25.223 & 0.003 \\
    &      & SCAD  & 1.061 & 0.138 & 17.006 & 0.000 & & 6.862 & 0.643 & 21.176 & 0.005 \\
    &      & ALASSO  & 1.361 & 0.139 &  17.893 & 0.000 & & 6.677 & 0.695 &  22.777 & 0.004 \\
    \\
    & 0.75 & LASSO & 2.451 & 0.686 & 18.456 & 0.000 & & 10.833 & 1.008 & 24.996 & 0.001 \\
    &      & SCAD  & 1.792 & 0.559 & 17.084 & 0.000 & & 8.789 & 0.899 & 21.088 & 0.005 \\
    &      & ALASSO  & 1.988 & 0.575 &  18.037 & 0.000 & & 8.994 & 0.951 &  21.050 & 0.003 \\
    \\
    & true $X$ & LASSO & 0.336 & 0.013 & 19.160 & 0.000 & & $-$ & $-$ & $-$ & $-$ \\
    &  & SCAD  & 0.331 & 0.009 & 17.796 & 0.000 & & $-$ & $-$ & $-$ & $-$\\
    &      & ALASSO  & 0.332 & 0.011 &  18.449 & 0.000 & & $-$ & $-$ & $-$ & $-$\\
     \hline     
40 (21) & 0.15 & LASSO & 0.810 & 0.044 & 25.768 & 0.000 & & 2.280 & 0.244 & 34.778 & 0.002\\
    &      & SCAD  & 0.632 & 0.029 & 22.626 & 0.000 & & 2.763 & 0.229 & 30.656 & 0.005 \\
    &      & ALASSO  & 0.774 & 0.033 & 22.578 & 0.000 & & 2.557 & 0.235 & 30.765 & 0.002 \\
    \\
   & 0.50 & LASSO & 1.352 & 0.205 & 24.470 & 0.000 & & 5.352 & 0.605 & 34.470 & 0.003 \\
    &      & SCAD  & 1.308 & 0.144 & 22.182 & 0.000 & & 4.308 & 0.544 & 30.182 & 0.005\\
    &      & ALASSO  & 1.255 & 0.123 & 21.430 & 0.000 & & 4.655 & 0.523 & 31.594 & 0.004\\
    \\
    & 0.75 & LASSO & 1.852 & 0.291 & 24.178 & 0.000 & & 7.752 & 0.691 & 34.178 & 0.003 \\
    &      & SCAD  & 1.744 & 0.259 & 22.766 & 0.000 & & 6.744 & 0.653 & 29.766 & 0.006 \\
    &      & ALASSO  & 1.754 & 0.264 & 22.000 & 0.000 & & 6.014 & 0.684 & 30.074 & 0.003 \\
    \\
    & true $X$ & LASSO & 0.324 & 0.013 & 23.251 & 0.000 & & $-$ & $-$ & $-$ & $-$ \\
    &  & SCAD  & 0.321 & 0.010 & 21.796 & 0.000 & & $-$ & $-$ & $-$ & $-$\\
    &      & ALASSO  & 0.342 & 0.012 &  21.598 & 0.000 & & $-$ & $-$ & $-$ & $-$\\
 \hline\hline
\end{tabular}
\end{table}

\clearpage

\begin{table}
       \huge
     \caption{Simulation result for Model 2 }\label{Simulation_B}

   \scriptsize

 \centering
   \renewcommand{\arraystretch}{0.9}
 \begin{tabular}{c c c c  c ccccc  ccc c c c c}

 \\
 \hline\hline
$p\ (q)$ & $\sigma_{\epsilon}$  & Method & \multicolumn{4}{c} { Result of proposed estimator $\widehat{\beta}$  } & & \multicolumn{4}{c} {Result of naive estimator $\widehat{\beta}_\text{naive}$ } \\ \cline{4-7} \cline{9-12}
 & & & $L_1$-norm & $L_2$-norm & \#S & \#FN
 & &  $L_1$-norm & $L_2$-norm & \#S & \#FN \\ 
 \hline 
30 (17)& 0.15 & LASSO & 1.150 & 0.078  & 20.197 & 0.000 & & 4.351 & 0.278  & 22.976 & 0.003\\
    &      & SCAD  &  0.879 & 0.046  & 19.142 & 0.000 & &  3.879 & 0.246  & 20.714 & 0.006\\
    &      & ALASSO  & 0.950 & 0.059 & 19.877 & 0.000 & & 4.295 & 0.269 & 20.970 & 0.004\\
    \\
   & 0.50 & LASSO &  1.161 & 0.081  & 20.298 & 0.000 & &  5.143 & 0.381  & 25.538 & 0.002\\
    &      & SCAD  & 0.869 & 0.046  & 19.980 & 0.000 & & 4.869 & 0.346  & 21.440 & 0.005\\
    &      & ALASSO  & 1.001 & 0.059 & 20.000 & 0.000 & & 4.303 & 0.358 & 22.037 & 0.003\\
    \\
    & 0.75 & LASSO & 1.692 & 0.165 & 23.486 & 0.000 & & 7.692 & 0.565 & 26.486 & 0.003\\
    &      & SCAD  & 1.508 & 0.130 & 22.042 & 0.000 & & 6.508 & 0.530 & 25.077 & 0.005\\
    &      & ALASSO  & 1.573 & 0.145 & 22.447 & 0.000 & & 6.713 & 0.542 & 25.387 & 0.003\\
    \\
    & true $X$ & LASSO & 0.892 & 0.049 & 19.246 & 0.000 & & $-$ & $-$ & $-$ & $-$\\
    &  & SCAD & 0.726 & 0.032 & 17.916 & 0.000 & & $-$ & $-$ & $-$ & $-$\\
    &      & ALASSO  & 0.759 & 0.040 & 18.334 & 0.000 & & $-$ & $-$ & $-$ & $-$\\
 \hline 
 40 (21)& 0.15 & LASSO & 0.954 & 0.054 & 26.094 & 0.000 & & 4.956 & 0.335 & 31.094 & 0.004\\
    &      & SCAD  & 0.584 & 0.026 & 22.162 & 0.000 & & 4.584 & 0.326 & 29.602 & 0.006\\
    &      & ALASSO  & 0.712 & 0.037 & 22.538 & 0.000 & & 4.782 & 0.334 & 29.548 & 0.003\\
    \\
    & 0.50 & LASSO & 1.365 & 0.093 &  26.702 & 0.000 & & 5.365 & 0.493 &  29.702 & 0.003\\
    &      & SCAD  & 1.058 & 0.058 & 22.622
 & 0.000 & & 5.058 & 0.458 & 27.622
 & 0.008\\
    &      & ALASSO  & 1.140 & 0.060 & 22.584 & 0.000 & & 5.400 & 0.460 & 27.433 & 0.005\\
    \\
    & 0.75 & LASSO & 1.305 & 0.099 &  26.356  & 0.000 & & 6.305 & 0.699 &  27.356  & 0.005\\
    &      & SCAD  & 1.182 & 0.087 & 22.802 & 0.000 & & 6.182 & 0.587 & 25.802 & 0.008\\
    &      & ALASSO  & 1.289 & 0.090  & 23.000 & 0.000 & & 5.289 & 0.612  & 26.013 & 0.006\\
    \\
    & true $X$ & LASSO & 0.883 & 0.033 & 24.870 & 0.000 & & $-$ & $-$ & $-$ & $-$\\
    &  & SCAD & 0.552 & 0.015 & 21.872 & 0.000 & & $-$ & $-$ & $-$ & $-$\\
    &      & ALASSO  & 0.648 & 0.026 & 22.483 & 0.000 & & $-$ & $-$ & $-$ & $-$\\
 \hline\hline
\end{tabular}
\end{table}

\clearpage

\begin{table}
       \huge
     \caption{Simulation result for Model 3 }\label{Simulation_C}

   \scriptsize

 \centering
   \renewcommand{\arraystretch}{0.9}
 \begin{tabular}{c c c c  c ccccc cccc ccc}

 \\
 \hline\hline
$p\ (q)$ & $\sigma_{\epsilon}$  & Method & \multicolumn{4}{c} { Result of proposed estimator $\widehat{\beta}$  } & & \multicolumn{4}{c} {Result of naive estimator $\widehat{\beta}_\text{naive}$ } \\ \cline{4-7} \cline{9-12}
 & & & $L_1$-norm & $L_2$-norm & \#S & \#FN
 & &  $L_1$-norm & $L_2$-norm & \#S & \#FN \\ 
 \hline 
30 (17) & 0.15 & LASSO & 0.716 & 0.065 & 20.092 & 0.000 & & 3.716 & 0.265 & 26.092 & 0.004 \\
    &      & SCAD  & 0.627 & 0.059 & 18.670 & 0.000 & & 3.627 & 0.259 & 24.670 & 0.005 \\
    &      & ALASSO  & 0.640 & 0.059 & 18.143 & 0.000 & & 3.640 & 0.248 & 25.179 & 0.004\\
    \\
   & 0.50 & LASSO & 0.884 & 0.085 & 21.532 & 0.000 & & 4.862 & 0.385 & 26.532 & 0.003 \\
    &      & SCAD  & 0.511 & 0.064 & 20.942 & 0.000 & & 4.511 & 0.320 & 24.142 & 0.006 \\
    &      & ALASSO  & 0.667 & 0.076 &  21.000 & 0.000 & & 4.567 & 0.366 &  24.110 & 0.004 \\
    \\
    & 0.75 & LASSO & 0.998 & 0.083 & 21.398 & 0.000 & & 5.998 & 0.483 & 27.398 & 0.003 \\
    &      & SCAD  & 0.609 & 0.064 & 20.786 & 0.000 & & 5.609 & 0.444 & 25.786 & 0.004 \\
    &      & ALASSO  & 0.771 & 0.075 &  20.000 & 0.000 & & 5.719 & 0.455 &  25.130 & 0.004\\
    \\
    & true $X$ & LASSO & 0.443 & 0.016 & 20.964 & 0.000 & & $-$ & $-$ & $-$ & $-$ \\
    &  & SCAD & 0.333 & 0.010 & 17.378 & 0.001 & &$-$ & $-$ & $-$ & $-$\\
    &      & ALASSO  & 0.334 & 0.011 &  18.000 & 0.000 & & $-$ & $-$ & $-$ & $-$ \\
    \hline     
40 (21) & 0.15 & LASSO & 1.290 & 0.101 & 23.450 & 0.000 & & 4.290 & 0.301 & 28.450 & 0.003 \\
    &      & SCAD  & 1.160 & 0.094 & 22.926 & 0.000 & & 4.160 & 0.294 & 27.926 & 0.003 \\
    &      & ALASSO  & 1.206 & 0.094 &  23.000 & 0.000 & & 4.156 & 0.284 &  26.112 & 0.004 \\
    \\
   & 0.50 & LASSO & 1.481 & 0.149 & 23.822 & 0.000 & & 4.481 & 0.349 & 30.822 & 0.003 \\
    &      & SCAD  & 1.061 & 0.138 & 22.006 & 0.000 & & 4.061 & 0.331 & 28.006 & 0.005 \\
    &      & ALASSO  & 1.161 & 0.139 &  22.000 & 0.001 & & 4.561 & 0.339 &  28.030 & 0.003 \\
    \\
    & 0.75 & LASSO & 1.551 & 0.186 & 24.456 & 0.000 & & 5.451 & 0.686 & 32.456 & 0.003 \\
    &      & SCAD  & 1.332 & 0.159 & 22.084 & 0.000 & & 4.792 & 0.599 & 28.084 & 0.005 \\
    &      & ALASSO & 1.488 & 0.151 &  22.000 & 0.000 & & 4.788 & 0.575 &  29.301 & 0.003 \\
    \\
    & true $X$ & LASSO & 0.936 & 0.073 & 23.160 & 0.000 & & $-$ & $-$ & $-$ & $-$ \\
    &  & SCAD  & 0.831 & 0.069 & 21.796 & 0.000 & & $-$ & $-$ & $-$ & $-$\\
    &      & ALASSO  & 0.842 & 0.072 &  22.000 & 0.000 & &$-$ & $-$ & $-$ & $-$\\
 \hline\hline
\end{tabular}
\end{table}

\clearpage

\begin{table}
       \huge
     \caption{Simulation result for Model 4 }\label{Simulation_D}

   \scriptsize

 \centering
   \renewcommand{\arraystretch}{0.9}
 \begin{tabular}{c c c c  c ccccc  ccc c c c c}

 \\
 \hline\hline
$p\ (q)$ & $\sigma_{\epsilon}$  & Method & \multicolumn{4}{c} { Result of proposed estimator $\widehat{\beta}$  } & & \multicolumn{4}{c} {Result of naive estimator $\widehat{\beta}_\text{naive}$ } \\ \cline{4-7} \cline{9-12}
 & & & $L_1$-norm & $L_2$-norm & \#S & \#FN
 & &  $L_1$-norm & $L_2$-norm & \#S & \#FN \\ 
 \hline 
30 (17)& 0.15 & LASSO &  0.689 & 0.037 & 20.190 & 0.000 & &  3.689 & 0.337 & 25.190 & 0.003 \\ 
    &      & SCAD  & 0.433 & 0.016 & 19.980 & 0.000 & & 3.433 & 0.316 & 23.980 & 0.005\\
    &      & ALASSO  & 0.571 & 0.029 & 20.000 & 0.000 & & 3.471 & 0.329 & 23.100 & 0.003\\ 
    \\
   & 0.50 & LASSO & 0.650 & 0.035 & 20.510 & 0.000 & & 4.650 & 0.455 & 27.510 & 0.003\\
    &      & SCAD  & 0.378 & 0.014  & 19.960 & 0.000 & & 4.378 & 0.414  & 24.960 & 0.007\\
    &      & ALASSO  & 0.407 & 0.027 & 20.030 & 0.000 & & 4.407 & 0.430 & 24.100 & 0.004\\
    \\
    & 0.75 & LASSO &  0.661 & 0.037  & 21.294 & 0.000 & &  5.174 & 0.507  & 26.294 & 0.004\\ 
    &      & SCAD  & 0.434 & 0.027  & 19.654 & 0.000 & & 4.934 & 0.470  & 23.654 & 0.006\\
    &      & ALASSO  & 0.450 & 0.037 & 20.100 & 0.000 & & 4.650 & 0.447 & 23.120 & 0.004\\
    \\
    & true $X$ & LASSO & 0.466 & 0.025 & 19.780 & 0.000 & & $-$ & $-$ & $-$ & $-$\\
    &  & SCAD & 0.312 & 0.019 & 17.850 & 0.000 & & $-$ & $-$ & $-$ & $-$\\
    &      & ALASSO  & 0.360 & 0.022 & 18.100 & 0.000 & &$-$ & $-$ & $-$ & $-$\\
    \hline     
40 (21)& 0.15 & LASSO & 1.150 & 0.078  & 24.976 & 0.000 & & 4.150 & 0.478  & 27.976 & 0.003\\
    &      & SCAD  &  0.879 & 0.046  & 22.142 & 0.000 & &  3.879 & 0.346  & 25.142 & 0.004\\
    &      & ALASSO  & 0.917 & 0.069 & 22.200 & 0.000 & & 3.950 & 0.379 & 25.600 & 0.003\\
    \\
   & 0.50 & LASSO &  1.161 & 0.081  & 23.298 & 0.000 & &  4.161 & 0.481  & 30.298 & 0.002\\
    &      & SCAD  & 0.869 & 0.046  & 21.980 & 0.000 & & 3.869 & 0.346  & 28.980 & 0.006\\
    &      & ALASSO  & 0.921 & 0.068 & 22.400 & 0.000 & & 3.903 & 0.389 & 29.400 & 0.003\\
    \\
    & 0.75 & LASSO & 1.192 & 0.085 & 24.486 & 0.000 & & 4.962 & 0.465 & 33.486 & 0.003\\
    &      & SCAD  & 0.858 & 0.043 & 21.042 & 0.000 & & 4.508 & 0.430 & 29.040 & 0.005\\
    &      & ALASSO  & 0.913 & 0.066 & 22.000 & 0.000 & & 4.713 & 0.442 & 29.000 & 0.004\\
    \\
    & true $X$ & LASSO & 0.892 & 0.049 & 23.246 & 0.000 & & $-$ & $-$ & $-$ & $-$\\
    &  & SCAD & 0.726 & 0.032 & 21.916 & 0.000 & & $-$ & $-$ & $-$ & $-$\\
    &      & ALASSO  & 0.759 & 0.040 & 22.100 & 0.000 & & $-$ & $-$ & $-$ & $-$\\
 \hline\hline
\end{tabular}
\end{table}

\clearpage

\begin{landscape} 
\begin{table}
       \huge
     \caption{Sensitivity analyses for analysis of WHAS500 data}\label{RDA_A}

   \small

 \centering
 
 \begin{tabular}{c c ccc c ccc c ccc c ccc}

 \\
 \hline\hline
& \text{Covariate}  & \multicolumn{3}{c} { $\sigma_\epsilon = 0.15$  } & & \multicolumn{3}{c} { $\sigma_\epsilon = 0.50$  } & & \multicolumn{3}{c} {$\sigma_\epsilon = 0.75$  } & & \multicolumn{3}{c} {naive estimator } \\ \cline{3-5} \cline{7-9} \cline{11-13} \cline{15-17}
& & LASSO & SCAD & ALASSO & & LASSO & SCAD & ALASSO & & LASSO & SCAD & ALASSO  & & LASSO & SCAD & ALASSO \\
 \hline 
& hr & 0.038 &  0 &  0 &  & 0.071 &  0 &  0  &  & 0.076 &  0 &  0 & & 0.026 & 0 & 0 \\
& sysbp & 0.120 &  0.040 &  0.036 &  & 0.118 & 0.025 &  0.019 &  & 0.121 &  0.028 &  0.027 & & 0.160 &0.150 &0.164\\
& diasbp & -0.064 & -0.096 & -0.126  & & -0.066 & -0.111 & -0.143 &  & -0.062 & -0.115 & -0.142 & &  -0.210 & -0.184 & -0.207\\
&  bmi & -0.142 & -0.174 & -0.204 &  & -0.139 & -0.185 & -0.216 &  & -0.140 & -0.193 & -0.220 & & -0.111 &-0.084 &-0.107\\
& cvd & 0 &  0 &  0 &  & 0 &  0 & 0 &  & 0& 0 &  0 &&0.001& 0 &0 \\
& afb & 0.120 &  0.039 &  0.036 &  & 0.118 &  0.025 &  0.019 & & 0.121 &   0.028 &  0.021 & & 0.183 & 0.157 & 0.180\\
& sho & 0 &  0 &  0  & & 0 &  0 & 0 & &  0 & 0 &  0 & &  0 & 0 &  0 \\
& age & 0.045 &  0 &  0 & & 0.046 &  0 & 0&  & 0.047 &0 &  0 && 0.096 & 0.079 & 0.095 \\
& gender & 0.045 &  0 &  0 &  & 0.060 &   0 &  0 &&  0.061 &  0 &  0 && 0.091 & 0.081 & 0.095\\
& chf & 0.086 &  0 &   0&  & 0.086 &  0 &  0 & & 0.088 &  0 &  0 & &  0 & 0 &  0 \\
& av3 & -0.132 & -0.164 & -0.194&  & -0.132 & -0.178 & -0.209 &&  -0.163 & -0.215 & -0.242 && -0.223 & -0.206 & -0.230 \\
& miord & -0.015 & -0.047 & -0.063&  & -0.017 & -0.063 & -0.094 &&  -0.024 & -0.076 & -0.104 && -0.103 & -0.077 & -0.010\\
& mitype & 0.575 &  0.495 &  0.514&  & 0.526 &  0.432 &  0.449&  & 0.496 &  0.4022 &  0.416 && 0.328 & 0.318 & 0.331\\
\hline
\#$S$ & &  11 & 7 & 7 &  & 11 & 7 & 7 &  & 11 & 7 & 7 && 11 & 9 & 9\\
 \hline\hline
\end{tabular}
\end{table}
\end{landscape}

\clearpage

\section*{References}

\refmark Bauldry, S., Bollen, K. A. and Adair, L. S. (2015) Evaluating measurement error in readings of blood pressure for adolescents and young adults. {\em Blood Pressure}, {24}, 96-102.

\refmark Buzas, J. F. (1998). Unbiased scores in proportional hazards regression with covariate measurement error. {\em Journal of Statistical Planning and Inference}, 67, 247-257.

\refmark Carroll, R. J., Ruppert, D., Stefanski, L. A., and Crainiceanu, C. M. (2006) {\em Measurement Error in Nonlinear Model}. Chapman \& Hall/CRC, New York.

\refmark Chen L.-P. (2018) Semiparametric estimation for the accelerated failure time model with length-biased sampling and
covariate measurement error. {\em Stat},7:e209. \\DOI: https://doi.org/10.1002/sta4.209

\refmark Chen, L.-P. and Yi, G. Y. (2019). Semiparametric methods for left-truncated and right-censored survival data with covariate measurement error. Submitted.

\refmark Chen, L.-P. (2019a) Pseudo likelihood estimation for the additive hazards model with data subject to left-truncation and right-censoring. {\em Statistics and Its Interface}, 12, 135–148.

\refmark Chen, L.-P. (2019b) Semiparametric estimation for cure survival model with left-truncated and right-censored data and covariate measurement error. arXiv:1812.11973. {\em Statistics and Probability Letters}, 154, 108547. DOI: 10.1016/j.spl.2019.06.023.

\refmark Chen, L.-P. (2019c). Statistical analysis with measurement error or
misclassification: Strategy, method and application. {\em Biometrics}, 75, 1045-1046.

\refmark Cook, J. R. and Stefaski, L. A. (1994) Simulation-Extrapolation Estimation in Parametric Measurement Error Models. {\em  Journal of the American Statistical Association}, 89, 1314 - 1328.

\refmark Fan, J. and Li, R. (2001) Variable selection via nonconcave penalized
likelihood and its oracle properties. {\em Journal of the American Statistical Association}, 96, 1348-1360.

\refmark Hosmer, D. W., Lemeshow, S. and May, S. (2008) {\em Applied Survival Analysis: Regression Modeling of Time to Event Data}. John Wiley and Sons Inc.

\refmark Hu, C. and Lin, D. Y. (2002). Cox regression with covariate measurement error. {\em Scandnavian Journal of Statistics}, 29, 637-655.

\refmark Huang, C. Y., Follmann, D. A., and Qin, J. (2012)  A maximum pseudo-profile likelihood estimator for the Cox model under length-biased sampling. {\em Biometrika}, 99, 199-210.

\refmark Huang, C. Y. and Qin, J. (2013) Semiparametric estimation for the additive hazards model with left-truncated and right-censored data. {\em Biometrika}, 100, 877-888.

\refmark Huang, Y. and Wang, C. Y. (2000). Cox regression with accurate covariates unascertainable: A nonparametric correction approach. {\em Journal of the American Statistical Association}, 95, 1209-1219.

\refmark Kalbfleisch, J. D. and Prentice, R. L. (2011) {\em The Statistical Analysis of Failure Time Data}. Wiley.

\refmark Lawless, J. F. ( 2003) {\em Statistical Models and Methods for Lifetime Data}. Wiley.

\refmark Lin, D. Y. and Wei, L. J. (1989) The robust inference for the Cox
proportional hazards model. {\em Journal of the American Statistical
Association}, 84, 1074-1078.

\refmark Lin, W. and Lv, J. (2013) High-dimensional additive hazards regression. {\em Journal of American Statistical Association}, 108, 247 - 264.

\refmark Nakamura, T. (1992). Proportional hazards model with covariates subject to measurement error. {\em Biometrics}, 48, 829-838.

\refmark Ning, J., Qin, J., and Shen, Y. (2014). Semiparametric accelerated failure time model for length-biased data with application to dementia study. {\em Statistica
Sinica}, 24, 313-333.

\refmark Qin, J. and Shen, Y. (2010) Statistical methods for analyzing right-censored length-biased data under Cox model. {\em Biometrics}, 66, 382-392.

\refmark Rothman K. J. (2008) BMI-related errors in the measurement of obesity. {\it International Journal of Obesity}, {32}, 56-59.

\refmark Shen, Y., Ning, J., and Qin, J. (2009). Analyzing length-biased data with semiparametric transformation and accelerated failure time models. {\em Journal of the
American Statistical Association}, 104, 1192-1202.

\refmark Su, Y. R. and Wang, J. L. (2012) Modeling left-truncated and right-censored survival data with longitudinal covariate. {\em The Annals of Statistics},  40, 1465-1488.

\refmark Tibshirani, R. (1996) Regression Shrinkage and Selection via the LASSO. {\em Journal of Royal Statistical Society, Series B}, 58, 267 - 288.

\refmark Wang, H., Li, R. and Tsai, C. (2007) Tuning parameter selectors for the smoothly clipped absolute deviation method. {\em Biometrika}, 94, 553 - 568.

\refmark Wang, M.-C. (1991). Nonparametric estimation from cross-sectional survival data. {\em Journal of the
American Statistical Association}, 86 130–143.

\refmark Xie, S. H., Wang, C. Y., and Prentice, R. L. (2001). A risk set calibration method for failure time regression by using a covariate reliability sample. {\em Journal of the Royal Statistical Society, Series B}, 63, 855-870.

\refmark Zou, H. (2006) The Adaptive Lasso and Its Oracle Properties. {\em Journal of the American Statistical Association}, 101, 1418 - 1429.

\end{document}